# TOWARDS AN INTEGRATED KNOWLEDGE MANAGEMENT AND INFORMATION AND COMMUNICATION TECHNOLOGY FRAMEWORK FOR IMPROVING DISASTER RESPONSE IN A DEVELOPING COUNTRY CONTEXT

Teurai Matekenya, Cape Peninsula University of Technology and University of Zimbabwe, teuraimtkn@gmail.com

Ephias Ruhode, Cape Peninsula University of Technology, RuhodeE@cput.ac.za

**Abstract:** This paper is part of an ongoing project that seeks to address a gap in disaster information coordination and collaboration in Zimbabwe. There is lack of coordinated information and knowledge in natural disaster and emergency situations in Zimbabwe. This results in weak collaboration links among the various organizations that respond to emergencies, leading to slow decision making processes and long response times. This negatively affects the affected communities, exacerbating poverty in Zimbabwe. This has been evidenced in the recent catastrophic cyclone Idai where many people were left dead, infrastructure destroyed and some people marooned. To address this, the research seeks to develop an integrated Knowledge Management and ICT framework that aid in coordination and collaboration among the various crisis responders. This will be achieved through a case study approach using Zimbabwe's Civil Protection Unit. PAR within DSRM will be used to gather data from CPU as well as with NGO respondents, traditional leaders and disaster response experts. Findings will be compared and contrasted with secondary data gathered in literature, this, with collected data will be used in developing a home grown coordination and collaboration solution. Qualitative approach to data collection will be adopted using interviews, visioning workshops and document analysis.

## 1. INTRODUCTION

Each year the world is struck by natural disasters which threaten human security and welfare (Oktari et al., 2020). Responding to such crisis involves a high demand for a critical mass of individuals and organisations who have different stakes in disaster recovery programs (LESLP, 2015). For effective disaster response, it is important that there be effective coordination and collaboration amongst these responders (Bjerge et al., 2016). This allows disaster responders to quickly and effectively respond to disaster, ultimately maximizing a nation's response capacity (Usada, 2017). However, coordination and inter-organisational collaboration in a disaster is complex (Kapucu, 2010) and responding effectively to disaster is a big challenge to most nations (Wang, 2013). According to Bjerge et al., (2016), a barrier to organized information sharing in disaster management is the availability of vast amounts of information which is sometimes not the relevant information that the stakeholder requires at that time. Bjerge et al., (2016) observed a gap in information coordination and sharing among responders which usually leads to overlapping initiatives, extensive resource mismanagement which ultimately leads to loss of lives and livelihoods (Provitolo, 2012). Alexander (2020) called for an urgent need for close coordination and collaboration among crises responders as crises such as the Corona virus, has pushed for effective information sharing within and across jurisdictional borders.

This paper is scoped empirically to examine the case of Zimbabwe, which has suffered natural disasters in recent years. In 2000, alongside Mozambique and South Africa, Zimbabwe was hit by Cyclone Eline. Reports indicated that 90 people died and over 250 000 marooned and approximately US$7.5 million in economic losses were experienced in Zimbabwe (Shumba, 2000). In March 2019, Zimbabwe, Mozambique and Malawi were hit by Cyclone Idai, which was characterised by heavy rains, mudslides and flooding. Cyclone Idai left hundreds of people dead, thousands marooned and infrastructure destroyed, with hundreds unaccounted for. The cyclone also left the governments overwhelmed with little resources to respond to the crisis. The Zimbabwean government activated its crisis coordinating organ, The Civil Protection Unit (CPU), to coordinate the emergency response. Unfortunately, CPU was incapacitated to coordinate such a catastrophic cyclone. A major challenge identified in Cyclone Idai was access to information to define the type of assistance required. A number of individuals, humanitarian actors, organisations that entered the affected provinces could not get timely information from CPU to assist them respond effectively to the emergency. It is therefore apparent that *there is a lack of coordinated information and knowledge in natural disaster and emergency situations in Zimbabwe. This results in weak collaboration links among the various organizations that respond to emergencies, leading to slow decision making processes and long response times. This negatively affects the affected communities, exacerbating poverty in the country.*

Under the above background, this paper seeks to answer the following question: *What are the key elements of a KM and ICT Framework that improves coordination and collaboration among emergency responders to natural disasters in Zimbabwe?*

The paper is structured as follows: the next section is a review of literature and theoretical framework. The section that follows presents the methodolody which involves data collection methods and techniques and the DSR used to design the rudimentary framework. The section that follows presents the expected outcome followed by a section on implications to theory and practice and lastly a summary.

## 2. LITERATURE REVIEW

### 2.1    Natural Disaster Management

Natural disasters destroy the people's common forms of survival such as health, food security, education as well as other related humanity aspects. No one organization, can succeed in Disaster Management (Singh, 2006). There is need for a number of organisations with different roles and expertise who partner and collaborate. The 1999 Odisha Cyclone in India resulted in increased Government-NGO collaboration (Singh, 2006) through regular coordination meetings, combined knowledge sharing, planning and vision construction (Ababe et al., 2008).

### 2.2    ICTs for Disaster Management

ICTs play a very pivotal role in information coordination and collaboration in all stages of the disaster management (Xue, 2017; Raymond et al., 2015). Despite their significant role, ICTs in disaster response has the following drawbacks among others: the disaster can result in power failure, hampering the use of ICT tools hence service delivery and communication systems can be oversubscribed resulting in communication difficulties. According to Comfort et al., (2004) there is need for creation of a flexible information infrastructure that manages the dynamic information exchange among the various emergency responders. For this system to be effective, it must ensure that relevant information gets to the right party timeously and in the right format to support prompt decision making (Stanton, 2016). According to Zhang et al, (2002) disaster situations require KM systems with collaborative technologies to ensure collaboration of multiple response organizations.

It is argued in this paper, that both KM and ICTs have the potential to provide for reliable interconnectedness of various organisations which are responsible for disaster response. KM focuses on systematic approach for searching and using knowledge with the overall aim of value making (Mráček & Mucha, 2015). Seneviratne et al., (2012) describe KM within the DM context as focusing on availing the correct knowledge to the right people in the exact place at the correct time.

**2.3 Theoretical Framework**

This paper recognizes the existence of many KM frameworks in literature. According to Holsapple and Joshi (1999), prominent frameworks include:

●      Framework of Knowledge Management Pillars;
●      Framework of Core Capabilities and Knowledge Building;
●      Model of Organizational Knowledge Management;
●      Framework of the Knowledge Organization; and
●      Framework of Knowledge Management Stages

While the rest of the frameworks focus on KM in one organisation, it is worthwhile to emphasise that the ICT/KM framework to be developed in this study can be expanded to conceptualise KM not just in one organisation. Thus it seeks to aid coordination and collaboration among disparate emergency responders and not just one organisation. In this paper, the Framework of KM Pillars shall be coalesced with the coordination theory (Malone, 1990) as shown in the next two subsections.

**2.3.1 Framework of Knowledge Management Pillars**

Stankosky et al., (2003) developed the Four Pillars of KM framework that organisations intending to embark on KM initiatives should consider for effective management of knowledge. The four pillars are Leadership, Organisation, Technology and Learning (Figure 1). The Leadership pillar focuses on strategically aligning KM initiatives with business objectives. The Organisation pillar focuses on redesigning and aligning of processes and procedures. The Technology pillar focuses on setting up an enabling technological infrastructure to support the KM initiative. The Learning pillar focuses on ways in which the organization creates a learning community (Stankosky et al., 2003).

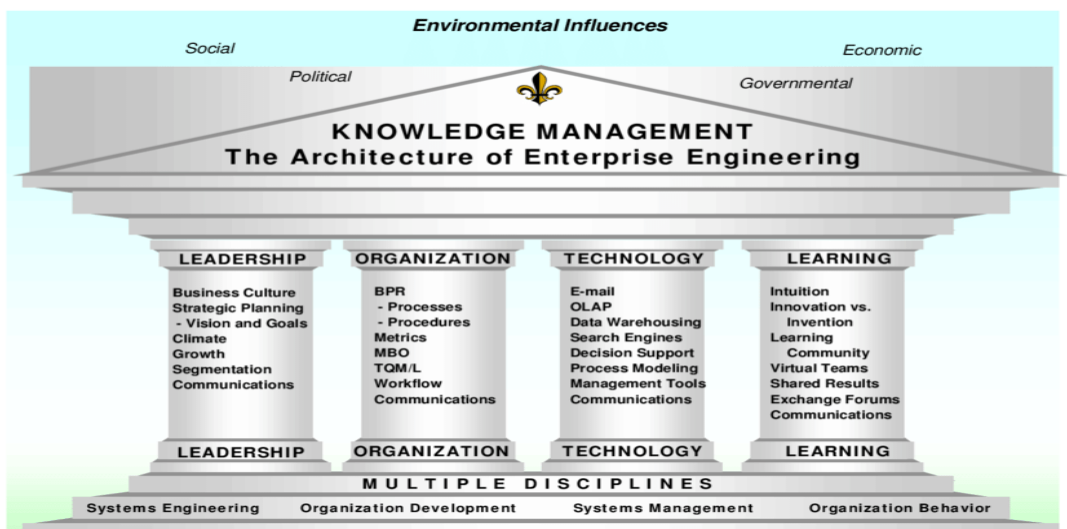

**Figure 1: The Four Pillars of Knowledge Management (Stankosky et al., 2003)**

### 2.3.2 Coordination Theory

Coordination is the act of working together harmoniously (Malone and Crowston, 1990). There is little known or coherent theory of coordination in the Information System domain. There are however numerous disjointed theories from various disciplines such as sociology, organizational theory, social psychology and many other disciplines. In all these various fields, there appears a common problem of coordination regardless of discipline which centers around how resources can be allocated effectively among different actors. Of importance to this study is "how information can be shared effectively with the different actors". According to Malone (1990), there are some processes underlying coordination (Table 1).

**Table 1: Processes underlying coordination (Malone, 1990)**

| Process Level | Components | Examples of generic process |
|---|---|---|
| Coordination | Goals, activities, actors, resources, interdependencies | Identifying goals, ordering activities, assigning activities to actors, allocating resources, synchronizing activities |
| Group decision making | Goals, actors, alternatives, evaluations, choices | Proposing alternatives, evaluating alternatives, making choices, (e.g., by authority, voting or consensus) |
| Communication | Senders, receivers, messages, languages | Establishing common language, selecting receiver (routing), transporting message (delivering) |
| Perception of common object | Actors, objects | Seeing same physical object, accessing shared databases. |

In the context of Zimbabwe, Table 2 shows the activities in the processes of disaster coordination as they relate to the Framework of Four Pillars of KM.

**Table 2: Coordination Theory Processes in the Four Pillars of Knowledge Management Leadership**

| Process Level | Leadership | Organisational | Technology | Learning |
|---|---|---|---|---|
| Coordination | Disaster coordination goals, disaster responders involved. Establish and implement KM strategy. Nourish the climate and culture for KM strategy implementation | Structures that manage actor interdependence. Operational processes aligned to the KM strategy. Redesigning the organization introducing change managers'/ KM champions on the organogram | For creation, capturing, storing, searching retrieving, sending only relevant information to each responder, sharing, fostering collaboration among responders, recommending solutions. | Increased internal and external communication, reports, developing institutional memory, lessons learnt database. Promoting cross-functional teams, creating a learning community. Interagency discussions |
| Group decision making | Mapping goals to activities, deciding on who does what, to whom and when. KM strategy formulation and Advocacy for humanitarian principles. | Reporting structure, nature and frequency of meetings. Compliance measures/codes of conduct, member complaints mechanism | Decision Support Systems. collaborative technologies; mobile communication applications, e.g. WhatsApp, Slack | Ensure the social processes of collaborating, sharing knowledge and building on each other's ideas are promoted to ensure tacit knowledge is shared. |
| Communication | Negotiating access to affected populations. Communication to the media | Mechanisms for collecting data from disaster population and for passing information to the various responders | Knowledge Dashboard, E-Mail, Business Intelligence, Analytical Processing. | Through Virtual Teams, sharing results, exchange forums |
| Perception of common object | Shared database for situational updates, collate needs assessments | Shared database for disaster information | Data Warehousing, Search Engines, Decision Support, Big Data Processing Tools. | Training of members/interagency through a standardized training curricula. |

CPU leadership should identify the disaster coordination goals, the disaster response actors, create a group of disaster response leadership with representative from NGOs or cluster leaders who will be involved in group decisions such as goal decomposition-who does what, to whom and where. This group should agree on the communication and KM strategy. Leadership should cultivate a culture of disaster information sharing so that responders have a common perception of the disaster situation.

**Organisation**

To support the leadership pillar, a number of organisational changes should be effected and managed. CPU's organisational structure, operational processes, staff, skills, systems should support the various disaster responders in accessing the right information at the right time in the right format. This should include setting up structures to manage actor interdependence, upgrade personnel skills to manage disaster information and knowledge, a KM champion on the organogram to drive the disaster KM initiative. Some group decisions should be made such as deciding and agreeing on the governance mechanism, reporting structure, frequency of meetings etc.

**Technology**

While organisational and cultural changes as highlighted in Pillars above are vital, a lack of a proper technology infrastructure leads to failure of KM initiative hence ineffective disaster coordination. There is need for IT tools that capture data, store, search and retrieve, send the relevant data to the various responders in a format appropriate to the responder, share the information and foster collaboration amongst the various responders, solve and provide recommendations to the responders for prompt decision making. The group should agree on the kind of tools and mechanism appropriate for effective collaboration.

**Learning**

For effective use of the technologies and implementation of the agreed KM strategy, CPU should create a conducive environment for learning. Organisational learning, using approaches such as promoting cross functional teams, building institutional memory, lessons learnt, group discussions, experience sharing as well as training should be adopted. The group should agree on the learning approach.

# 3. METHODOLOGY

A case study approach using Zimbabwe's disaster coordinating agency-(CPU) will be adopted as it helps in acquiring in-depth understanding of how CPU carries out the coordinating role.Cohen et al., (2007) argues that case studies provide distinctive examples of genuine people in real situations. Design Science Research (DSR) will be most appropriate in this study as it is a practical research method that produces a technology based solution. According to Baskerville et al (2015), one mandate of DSR is to produce practical solutions to teething problems by bringing change through improving existing systems. According to Peffers et al (2007), DSR includes the following six steps: 1. Problem identification; 2. Defining the research objectives; 3. Designing and developing the artefacts; 4. Demonstrating; 5. evaluating the solution by matching the objectives and the observed results from the use of the artefacts and 6. Communicating the problem, the artefact, its usefulness and effectiveness to other practitioners and researchers.

## 3.1    Data collection Methods and techniques

In order to answer the main research question, two sub questions will be asked. These are "What are the current coordination and collaboration practices employed by CPU? and " What are the emergency responders' service expectations from CPU? PAR will be used within the DSRM to get an understanding of Zimbabwe's CPU. Futures visioning workshopping shall be the dominant method for data collection. Visioning is a participatory approach that brings a group of stakeholders together and supports them in developing a shared vision of the future. To answer the three questions, data collection will be done in stages as follows:

**Stage 1: Current realities assessment**

This involves collecting data from CPU through interviews with Key Informants (KI), document analysis and extensive literature search to understand how other countries are coordinating emergency response. The aim of this stage will be to better understand CPU's current coordination mechanism. Specific information to be collected include CPU's vision, workflow, current technologies and others as shown in Table 3 below. This will be the first step towards PAR, involving Ministry KI in defining the problem so as to solve it (Avison et al., 2001). Findings from this will be used to feed into stage 2.

**Stage 2: Future's visioning**

Data will be collected through a future's visioning workshop that will be attended by respondents from NGO and CPU. Respondents will highlight aspects of the past and present they think are the most important elements to include in a future vision of CPU. The purpose of this exercise will be to identify what CPU's stakeholders consider as important for effective coordination and collaboration. The goal of this phase will be to arrive at a vision that reflects the thinking of diverse stakeholders and not majority opinion.

**Stage 3: Identifying mechanisms to support the envisioned future**

Participants will form into new smaller groups and members will be asked to brainstorm the changes that should take place at CPU inorder to ensure effective coordination and collaboration. The facilitator will introduce the 7s Mckinsey, a model for organisational effectiveness to assist the participants in coming up with changes/adjustments that should support the envisioned future. The model postulates that there are seven factors that need to be in alignment for an organisation to be successful. The identified changes will be used to develop the integrated ICT/KM Framework.

**3.2 Data analysis**

Data collected from the three stages will be collated and analysed thematically and manually. Thematic analysis identifies, analyses and reports themes/patterns within data (Braun & Clarke, 2006). The four pillars of KM coalesced with the coordination theory (Leadership, Organisation, Technology and Learning) will form the themes. This will be used in the development of the integrated KM and ICT framework that will ensure effective coordination and collaboration among the emergency responders. Table 3 shows a summary of data collection and analysis.

|  | **Stage 1: Current Realities assessment** | **Stage 2: Futures' visioning** | **Stage 3: Identifying mechanisms to support the envisioned future** |
|---|---|---|---|
| **Research question to be answered** | What are the current coordination mechanisms and collaboration practices employed by CPU? | What are the emergency responders' service expectations from CPU? | What are the key elements of a KM and ICT Framework that improves coordination and collaboration among emergency responders to natural disaster in Zimbabwe? |
| **Method for data collection** | Key informant interviews with CPU representatives<br><br>Document analysis-analysing Ministry documents as well as extensive literature search | Through a future's visioning workshop | Through the future's visioning workshop<br>7s Mckinsey model to assist the participants in coming up with changes/adjustments that should support the envisioned future |
| **Information to be collected** | Coordinating agency's vision, workflow, its challenges in coordinating emergency response, current technologies facilitating coordination and collaboration, the strength, weaknesses, opportunities and threats as well as agency's needs.<br><br>Technical questions : current system architecture, type of data being collected, volumes of data and how they are using the current collected data. | Aspects of the past and present respondents think are most important to include in a future vision of the coordinating organ. | Changes that should take place at CPU in order to get to the future. This will be grouped according to the four pillars of KM [Leadership, Organisation, Technology and Learning] |
| **Data analysis** | Data collected will be analysed thematically according to the four pillars of KM | Data will be analysed by the whole group as the workshop goes and the researcher will group the findings according to the four pillars of KM | Data collected from the three stages at each of the cases will be collated and analysed thematically and manually according to the four pillars |

**Table 3: Data collection and analysis summary**

### 3.3    Framework development and validation

Based on the extensive literature search and data collected, the key elements to the framework will be identified. The proposed framework will be validated by demonstrating its viability and suitability to the identified problem through a two-day workshop. The workshop participants will include those that would have been engaged in problem awareness stage of the DSR approach. The participants will discuss the applicability of the tools, techniques and procedures proposed in the framework and evaluate the extent to which the framework addresses the problem.

## 4. EXPECTED OUTCOME: DISASTER RESPONSE FRAMEWORK (DRF)

The DSR iterative process will lead to a participatory design outcome, which is the disaster response framework (DRF) which will be informed by ICT and KM concepts. The envisaged DRF will be guided by the four pillars of KM which in this research, have been adapted and applied to the disaster response empirical context (Figure 2).

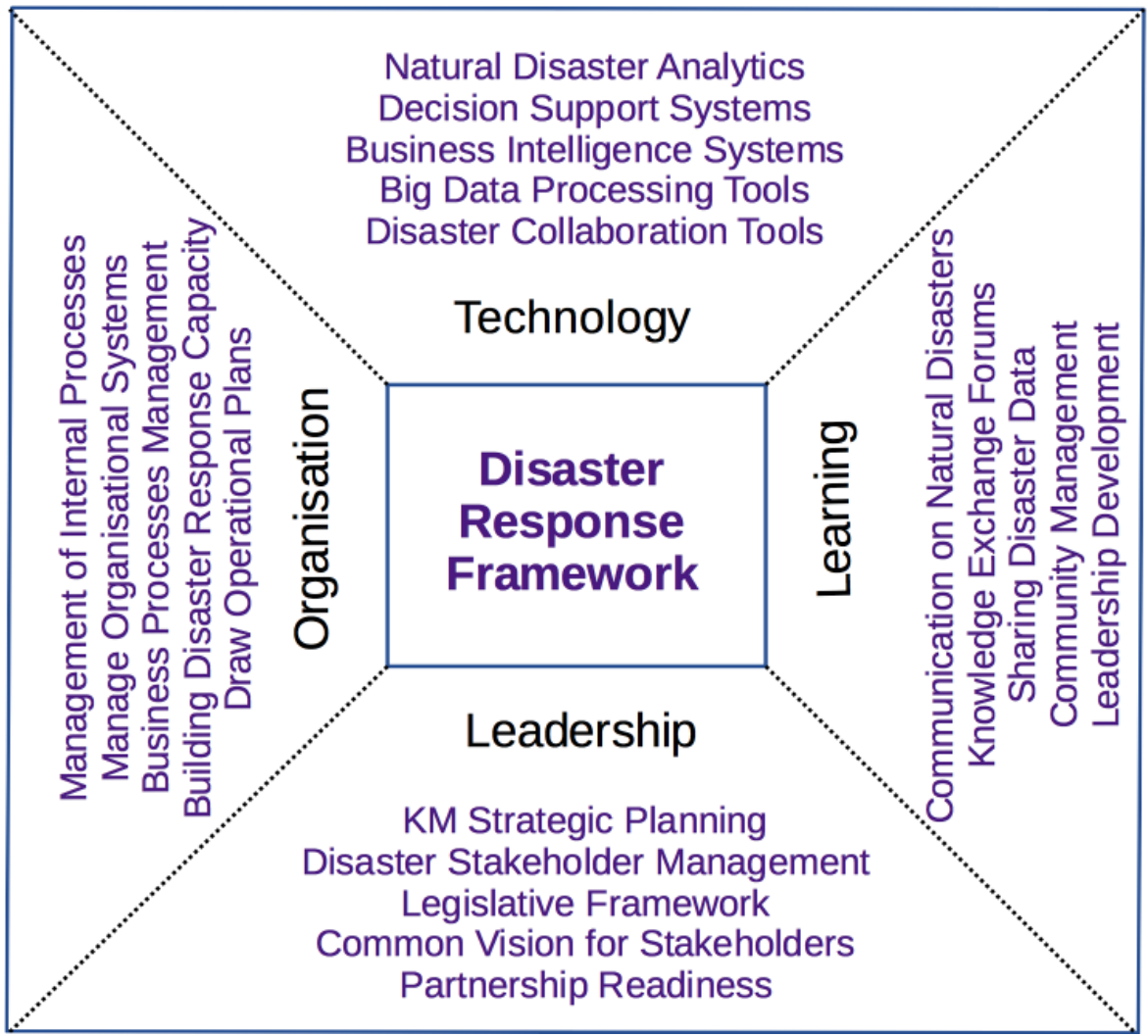

**Figure 2: Envisaged Disaster Response Framework (DRF)**

## 5. IMPLICATIONS TO THEORY AND PRACTICE

*Theoretically* : Findings from the empirical study will be used to extend and complements earlier theories by providing a novel Zimbabwean country perspective on disaster coordination and collaboration.

*Practically* **:** An implementable integrated KM and ICT framework for improving coordination and collaboration will be developed.

*Methodologically* : This research uses a combination of DSR, PAR, collaboration theory and Framework of KM Pillars in exploring the current coordination and collaboration problems at CPU and in devising the solution. This is the first research in developing country context that has used these methodologies specifically in improving emergency response.

## 6. SUMMARY

The paper has presented part of an ogoing project that seeks to address a gap in disaster information coordination and collaboration in Zimbabwe. Literature has revealed that ICTs and KM has the potential of addressing the gap hence the review of theories on KM and coordination which will be coalesced to develop the DRF. The paper proposes the use of PAR's visioning workshopping within the DSRM in coming up with the envisioned DRF.

# REFERENCES AND CITATIONS

Ababe, D., Cultis, A., Aklilu, Y., Mekonnon, G., and Ghebrechirstos, Y. (2008). Impact of a commercial destocking relief intervention in Moyale District, Southern Ethiopia. Overseas Development Institute. Blackwell Publishing.

Alexander, D.J (2020). Information sharing more critical than ever amid the Coronavirus.

Retrieved from https://www.govtech.com/em/safety/Information-Sharing-More-Critical-Than-Ever-amid-the-Coronavirus.html. [Accessed 13 March 2020].

Avison, D., Baskerville, R., and Myers, M. (2001). Controlling action research projects. IT & People. 14. 28-45. 10.1108/09593840110384762.

Baskerville, R,L.; Kaul, M; and Storey, V,C., (2015). "Genres of Inquiry in Design-Science

Research: Justification and Evaluation of Knowledge Production," MIS Quarterly, (39: 3) pp.541-564.

Bjerge, B., Clark, N., Fisker P., and Raju, E., (2016). Technology and Information Sharing in Disaster Relief. PLoS ONE 11(9)

Braun, V. & Clarke, V. (2006) Using Thematic Analysis in Psychology. Qualitative Research in Psychology, 3, 2, 77-101.

Cohen, L., Manion, L., & Morrison, K. (2007). Research Methods in Education. London: Routledge.

Comfort, L,K., Ko, K and Zagorecki, A., (2004). "Coordination in Rapidly Evolving Disaster Response Systems: The Role of Information". American Behavioural Scientist 48: 295.

Holsapple, C.W. and Joshi, K.D., (1999). Description and Analysis of Existing Knowledge Management Frameworks. Conference Paper · February 1999

Kapucu, N., Arslan, T and Collins, Lloyd, M., (2010) "Examining Intergovernmental and Interorganizational Response to Catastrophic Disasters: Toward a Network-Centered Approach." Administration & Society 42: 222.

London Emergency Services Liaison Panel. (2015). Major incident procedure manual (9.3 ed.).

Malone, T. W.1990. Organizing information processing systems: Parallels between organizations and computer systems. In Zachary, W., Robertson, S., Black, J. (Ed.), Cognition, Computation. and Cooperation (pp. 56-83). Norwood, NJ: Ablex

Malone, T.W and Crowston, K. (1990). What is coordination theory and how can it help design cooperative work systems?. CSCW '90: Proceedings of the 1990 ACM conference on Computer-supported cooperative workSeptember 1990 Pages 357–370https://doi.org/10.1145/99332.99367

Mráček, P., and Mucha, M., (2015). The Use of Knowledge Management in Marketing Communication of Small and Medium-Sized Companies/ Procedia - Social and Behavioral Sciences 175 185 – 192

Oktari R, S., Munadi, K. Idroes, R., Sofyan H., (2020).Knowledge management practices in disaster management: Systematic review, International Journal of Disaster Risk Reduction, Volume 51, 101881, ISSN 2212-4209

Peffers, K., Tuunanen, T., Rothenberger, M.A, and Chatterjee, S. (2007). A Design Science Research Methodology for Information Systems Research. Journal of Management Information Systems, Volume 24 Issue 3, Winter 2007-8, pp. 45-78

Provitolo, D., (2012). The Contribution of Science and Technology to Meeting the Challenge of Risk and Disaster Reduction in Developing Countries: From Concrete Examples to the Proposal of a Conceptual Model of "Resiliency Vulnerability". Technologies and Innovations for Development, Springer: 131-153

Raymond, N., Card, A., Brittany, L., Achkar, Z.A., (2015). What is 'Humanitarian Communication'? Towards Standard Definitions and Protections for the Humanitarian Use of ICTs. European Interagency Security Forum.

Seneviratne, S.I., N. Nicholls, D. Easterling, C.M. Goodess, S. Kanae, J. Kossin, Y. Luo, J.

Marengo, K. McInnes, M. Rahimi, M. Reichstein, A. Sorteberg, C. Vera, and X. Zhang, (2012): Changes in climate extremes and their impacts on the natural physical environment. In:

Managing the Risks of Extreme Events and Disasters to Advance Climate Change Adaptation [Field, C.B., V. Barros, T.F. Stocker, D. Qin, D.J. Dokken, K.L. Ebi, M.D. Mastrandrea, K.J. Mach, G.-K. Plattner, S.K. Allen, M. Tignor, and P.M. Midgley (eds.)]. A Special Report of Working Groups I and II of the Intergovernmental Panel on Climate Change (IPCC). Cambridge University Press, Cambridge, UK, and New York, NY, USA, pp. 109-230.

Shumba, O. (2000) An assessment of NGO Disaster Mitigation and Preparedness Activities in Zimbabwe: Country Survey. SAFIRE, Harare

Singh, M. (2006). Identifying and assessing drought hazard and risk in Africa. Regional conference on insurance and reinsurance for natural catastrophe risk in Africa. Casablanca, Morocco, November 14/12/2005.

Stankosky, M. F., Calabrese, F., & Baldanza, C. (2003). A systems approach to engineering aknowledge management sys-tem. Washington, DC: Management Concepts Press.

Stanton, N. A.,( 2016). Distributed situation awareness. Theoretical Issues in Ergonomics Science,17(1), 1–7.

Usuda, Y., (2017). "Shared Information Platform for Disaster Management with integrating government agencies in Japan," Nihon Jishinkogakkai Taikai Yokoshu (Proc. of the Conf. of Japan Association for Earthquake Engineering), CD-ROM, (in Japanese).

Wang, J.J., (2013). Post-disaster cross-nation mutual aid in natural hazards: case analysis from sociology of disaster and disaster politics perspectives. Natural Hazards. 66(2):413–38.

Xue, (2017).  A Literature Review on Knowledge Management in Organizations. Research in Business and   Management ISSN 2330-8362 2017, Vol. 4, No. 1

Zhang, D; Zhou, L; Jr,  Nunamaker J, F (2002) . A Knowledge Management Framework for the Support of Decision Making in Humanitarian Assistance / Disaster Relief. 370-385